\documentclass[journal,draftcls,onecolumn]{IEEEtran}
\usepackage{amsmath, amssymb}
\usepackage{graphicx}
\usepackage{algorithm}
\usepackage{algorithmic}
\usepackage{subfig}

\ifCLASSINFOpdf
\else
\fi
\begin{document}
\title{Multichannel RSS-based Device-Free Localization with Wireless Sensor Network}
\author{Zhenghuan Wang, Heng Liu, Shengxin Xu, Xiangyuan Bu , Jianping An,~\IEEEmembership{Member,~IEEE}% <-this % stops a space
\thanks{This work was supported by the National Science Foundation of China under Grant 61101129, 61227001.

The authors are with the School of Information and Electronics, Beijing Institute of Technology, Beijing 100081, China (e-mail: wangzhenghuan@bit.edu.cn; lhengzzt@bit.edu.cn; bxy@bit.edu.cn;an@bit.edu.cn).}}
\maketitle

\begin{abstract}
RSS-based device-free localization (DFL) is a very promising technique which allows localizing the target without attaching any electronic tags in wireless environments. In cluttered indoor environments, the performance of DFL degrades due to multipath interference. In this paper, we propose a multichannel obstructed link detection method based on the RSS variation on difference channels. Multichannel detection is proved to be very effective in multipath environments compared to the single channel detection. We also propose a new localization method termed as robust weighted least square (RWLS) method. RWLS first use spatial property to eliminate the interference links and then employ WLS method to localize the target. Since the spatial detection relies on the unknown position of the target. A coarse position estimation of target is also presented. RWLS is robust to interference links and has low computation complexity. Results from real experiments verify the effectiveness of the proposed method.
\end{abstract}


% \fi
\IEEEpeerreviewmaketitle

\section{Introduction}
\IEEEPARstart{D}{evice}-free localization (DFL) is an emerging technology for localizing target without attaching any electronic tags in the monitored area. It can be widely used for applications in home security, emergency response, military operations and other potential applications. Most DFL methods employ video camera, infrared or acoustic sensors and radar to passively detect and localize the target. However, these methods are limited by the capability of penetration or cost. In recent years, RSS-based DFL has attract lots of attention since the RSS can be acquired in most of the wireless equipment, for instance, WiFi access points or wireless sensor nodes [1-2]. Therefore, RSS based DFL could be easily extended to current wireless network without extra hardware. Moreover, this technology had advantages that radio signals can penetrate walls or other non-metallic structures [4]. Currently, RSS-based DFL has been successfully applied in monitoring [3-10], simultaneous localization and mapping (SLAM) [11-13], roadside surveillance [14], through-wall sensing [15-16], life detection [17] and fall detection for elder persons [18].

RSS-based on DFL exploits the RSS variation caused by the presence of the target. For instance, when the target enters into the monitored area, the target will absorb, reflect or scatter the radio signals. Most but not all DFL methods mainly utilize the attenuation of RSS when a link is obstructed by the target because obstructed link can offer useful position of the target [4]. Hence DFL can be divided into two steps: obstructed links detection and localization.

Most existing methods simply detect obstructed links based on the attenuation of RSS under a single channel. It is appropriate for outdoor environment where LOS signal of a link is dominant. However, for indoor environments, due to the multipath, the RSS variation of a link is unpredictable, leading to large detection error under a single channel. To overcome this problem, Wilson [15] proposed a variance based method to detect affected link. He observed when a person moves around a link, the RSS of the link will change rapidly. But this method is only effective to detect motion target. Kaltiokallio [19-20] proposed a channel diversity method which weights the RSS on each channel by fade level. The author defined the fade level by the difference of measured RSS and predicted RSS according to path loss model. However, for indoor environment, it's inappropriate to employ the same path loss model for every links because the propagation path each link is different. Zanella [21] also noticed that averaging RSS on multiple channels can greatly improve the ranging performance.

For localization method, Wilson proposed a technology termed as radio tomographic imaging (RTI) [3-20] which generates an attenuation image about the monitored area. Hence the localization accuracy is limited by the size of the grid in the image. Li [22-23] established a nonlinear measurement model relating to target's position and use particle filtering [22-25] to track the target.  Savazzi [26] also derived a similar model based on the diffraction theory. The accuracy of measurement models degrades in cluttered environments and the computational complexity of particle filtering is too high.

In this paper, we proposed a multichannel RSS-based obstructed link detection. Due to the fact that RSS on different channels varies substantially, we use the variance of RSS on the channels to estimation the attenuation of the LOS path and then use the attenuation estimation to detect the obstructed links. Compared to obstructed detection under a single channel which is vulnerable to multipath, multichannel detection performs well even in cluttered indoor environments. We also propose a weighted least square (WLS) localization method. WLS does not require the specific model between RSS change and the position of the target, and the computation burden  of WLS is very small. However, like most localization methods except RTI, WLS is sensitive to the non-obstructed links caused by false alarm and multipath interference. To solve the problem, we proposed a robust WLS (RWLS) localization method. RWLS first uses spatial property of obstructed link to eliminate the non-obstructed links. Spatial detection is based on the fact that the RSS of LOS path signal attenuates only when the target is close the link. Hence, spatial detection depends on the unknown position of target. To this end, we first use a simple method, which is similar to RTI but without regularization, to get coarse position estimation target and then use spatial property to discard the interference links. Finally we employ WLS to enhance the localization accuracy. We also conduct experiment in indoor environment to investigate the effectiveness of our method. The experiment results show the performance of detection and localization is greatly enhanced.

The rest of the paper is organized as follows. In Section 2, we formulate the obstructed links detection method using multichannel RSS. In Section 3, we propose the RWLS method to localize the target. Section 4 describes the experiment platform and Section 4 presents the results of our experiment. We conclude the paper in Section 5.

\section{Obstructed Link Detection }
Consider a monitored area consisting of $K$sensors with known position $\left( {{x}_{i}},{{y}_{i}} \right),i=1,...,K$, as shown in Fig.1. The nodes are placed around the perimeter of the monitored area with same height off the ground. The monitored area is rectangular with size $\left[ {{x}_{\min }},{{x}_{\max }} \right]\times \left[ {{y}_{\min }},{{y}_{\max }} \right]$. A pair of sensors can constitute a unique link and $N$ fully connected sensors can constitute $L=K\left( K-1 \right)/2$ links. The sensors can operate on $C$ different channels and measure the RSS of $L$ links. We denote the RSS of link $l$ on channel $c$ as ${{\bar{P}}_{l,c}},l=1,2,...,L,c=1,2,...,C$ when the monitored area is absent of target, where the unit of ${{\bar{P}}_{l,c}}$ is mW. When the target with coordinate $\left( x,y \right)$ is present, the target will obstruct, scatter or reflect radio signals, resulting in the RSS of link $l$ denoted as ${{P}_{l,c}}$ changed. DFL enable people to localize the target by means of the variation of RSS.
\subsection{Obstructed Links Detection}

\begin{figure}[!t]
\centering
\includegraphics[width=2.5in]{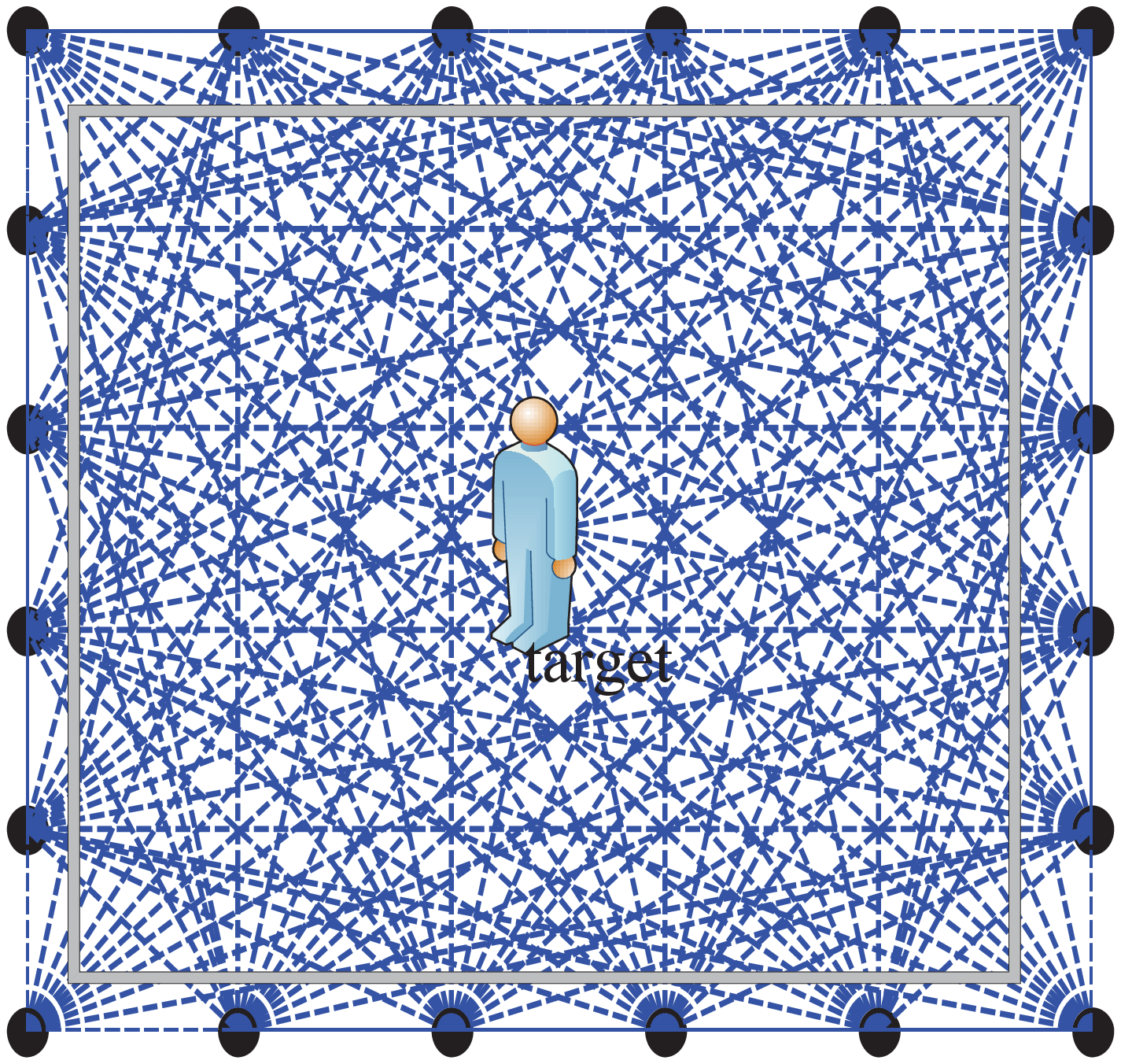}
\caption{Monitored area constituted by wireless sensors.}
\label{fig_sim}
\end{figure}

In DFL, the first step is to detect the links affected by the target. Here, the affected links should be the links obstructed by the target because only those links can provide useful information about the target.  In outdoor environments where LOS path is dominant, obstruction can be simply detected according to attenuation since the RSS of a link is severely attenuated when the link is blocked by the target.

\begin{figure}[!t]
\centering
\includegraphics[width=2.5in]{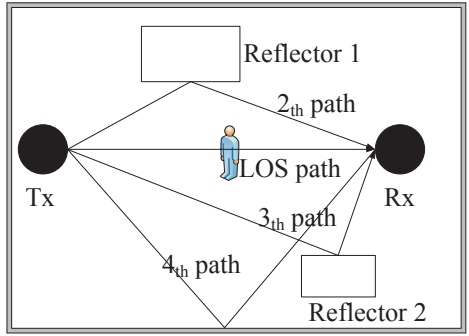}
\caption{Illustration of a multipath environment.}
\label{fig_sim}
\end{figure}

For indoor environments which is rich in multipath, as illustrated in Fig.2. There are many reflectors in the environment, allowing signals can propagate from transmitter to receiver via multiple paths apart from LOS path. Therefore, the variation of RSS links can be caused by target obstruction or multipath interference. In single channel case, it's difficult to detect the obstruction according to the variation of RSS, especially when LOS path is not dominant. In fact, the change of RSS is unpredictable because the received signal is phasor sum of the all paths. If the LOS path is constructive to the received signal, the RSS tends to be attenuated when LOS path is obstructed. On the other hand, when the LOS path signal is deconstructive to the received signal, the RSS will be enhanced.  Hence, it is unreliable for link obstruction detection using a single channel RSS in indoor environment.

In multichannel case, the link detection accuracy can be enhanced if the RSS variation on different channels is considered. We suppose there are $P$ paths for a link and signal of the ${{i}_{th}}$ path on $c$ channel is denoted by ${{A}_{i,c}}{{e}^{j{{\theta }_{i,c}}}}$. According to path loss model [29], that is,
\begin{equation}
A_{i,c}^{2}\propto \frac{1}{4\pi f_{c}^{2}}
\end{equation}
where ${{f}_{c}}$ is center frequency of transmitted signal on $c$ channel. In low-cost DFL, the transmitted signal is narrowband, that is, the bandwidth of signal is much lower than the center frequency. For example, most wireless sensors are compatible with IEEE802.15.4 standard  which specifies 16 channels on 2.4G frequency band with center frequency range from 2.405GHz to 2.480GHz [28]. The ratio between bandwidth and the center frequency is only 0.03. Based on the fact, the amplitude of ${{i}_{th}}$ path signal can be regarded as identical on different channels, i.e., ${{A}_{i,1}}={{A}_{i,2}}=,...,{{A}_{i,P}}$. And the phase is assumed to be random due to the nonlinear phase response of the channels. Therefore, the received signal of each channel can be considered a realization of the signal with random phase:
\begin{equation}
r=\sum\limits_{i=1}^{P}{{{A}_{i}}{{e}^{j{{\theta }_{i}}}}}
\end{equation}

The power of the received signal is
\begin{equation}
{{P}_{1}}={{\left| \sum\limits_{i=1}^{P}{{{A}_{i}}{{e}^{j{{\theta }_{i}}}}} \right|}^{2}}=\sum\limits_{i=1}^{P}{A_{i}^{2}+2}\sum\limits_{j=i+1}^{P}{\sum\limits_{i=1}^{P}{{{A}_{i}}{{A}_{j}}\cos \left( {{\theta }_{i}}-{{\theta }_{j}} \right)}}
\end{equation}
${{P}_{1}}$ is obtained when the target is absent in the monitored area. The mean of the power ${{P}_{1}}$ is
\begin{equation}
\begin{aligned}
  & E\left( {{P}_{1}} \right)=\sum\limits_{i=1}^{P}{A_{i}^{2}+2}\sum\limits_{j=i+1}^{P}{\sum\limits_{i=1}^{P}{{{A}_{i}}{{A}_{j}}E\left( \cos \left( {{\theta }_{i}}-{{\theta }_{j}} \right) \right)}} \\
 & =A_{1}^{2}+\sum\limits_{i=2}^{P}{A_{i}^{2}} \\
\end{aligned}
\end{equation}

We can see the average power is the power sum of all paths. Suppose $\left( {{\theta }_{i}}-{{\theta }_{j}} \right)$ is uniformly distributed in the interval $[0,2\pi ]$.The variance of signal power is

\begin{equation}
\operatorname{var}\left( {{P}_{1}} \right)=2A_{1}^{2}\sum\limits_{i=1}^{P}{A_{2}^{2}}+2\sum\limits_{j=i+1}^{P}{\sum\limits_{i=2}^{P}{A_{i}^{2}A_{j}^{2}}}
\end{equation}

Suppose a target obstructs the LOS path signal, causing the LOS path attenuated by $\gamma$ dB. Then the power of received signal becomes

\begin{equation}
\begin{aligned}
  & {{P}_{2}}={{\left| {{A}_{1}}{{10}^{-\gamma /20}}{{e}^{j{{\theta }_{1}}}}+\sum\limits_{i=2}^{P}{{{A}_{i}}{{e}^{j{{\theta }_{i}}}}} \right|}^{2}}=\sum\limits_{i=1}^{P}{{{10}^{-\gamma /10}}A_{i}^{2}}+ \\
 & \sum\limits_{i=1}^{P}{{{10}^{-\gamma /20}}{{A}_{i}}{{A}_{j}}\cos \left( {{\theta }_{i}}-{{\theta }_{j}} \right)}+2\sum\limits_{j=i+1}^{P}{\sum\limits_{i=2}^{P}{{{A}_{i}}{{A}_{j}}\cos \left( {{\theta }_{i}}-{{\theta }_{j}} \right)}} \\
\end{aligned}
\end{equation}

The mean and variance of the obstructed signal's power are given by
\begin{equation}
\begin{aligned}
  & E\left( {{P}_{1}} \right)=={{10}^{-\gamma /10}}A_{1}^{2}+\sum\limits_{i=2}^{P}{A_{i}^{2}} \\
 & \operatorname{var}\left( {{P}_{1}} \right)=\operatorname{var}\left( 2\sum\limits_{j=i+1}^{P}{\sum\limits_{i=1}^{P}{{{A}_{i}}{{A}_{j}}\cos \left( {{\theta }_{i}}-{{\theta }_{j}} \right)}} \right) \\
 & ={{10}^{-\gamma /10}}A_{1}^{2}\sum\limits_{i=1}^{P}{2A_{2}^{2}}+2\sum\limits_{j=i+1}^{P}{\sum\limits_{i=2}^{P}{A_{i}^{2}A_{j}^{2}}} \\
\end{aligned}
\end{equation}

The 1 order and 2 order moment based estimators about $\gamma $ are

\begin{equation}
\begin{aligned}
  & \hat{\gamma }=10{{\log }_{10}}\frac{E\left( {{P}_{1}} \right)}{E\left( {{P}_{2}} \right)}=10{{\log }_{10}}\frac{A_{1}^{2}+\sum\nolimits_{i=2}^{P}{A_{i}^{2}}}{{{10}^{-\gamma /10}}A_{1}^{2}+\sum\nolimits_{i=2}^{P}{A_{i}^{2}}} \\
 & \hat{\gamma }=10{{\log }_{10}}\frac{\operatorname{var}\left( {{P}_{1}} \right)}{\operatorname{var}\left( {{P}_{2}} \right)}=10{{\log }_{10}}\frac{2A_{1}^{2}\sum\nolimits_{i=2}^{P}{A_{i}^{2}}+2\sum\nolimits_{j=i+1}^{P}{\sum\nolimits_{i=2}^{P}{A_{i}^{2}A_{j}^{2}}}}{{{10}^{-\gamma /10}}2A_{1}^{2}\sum\nolimits_{i=2}^{P}{A_{i}^{2}}+2\sum\nolimits_{j=i+1}^{P}{\sum\nolimits_{i=2}^{P}{A_{i}^{2}A_{j}^{2}}}} \\
\end{aligned}
\end{equation}

Both mean and variance based estimator can be served for attenuation estimation. It's obvious that when $P=1$, meaning multipath is absent, $E\left( {\hat{\gamma }} \right)=\gamma$. But when $P>1$, mean based estimator is always  biased . For variance based estimator, when $P=2$, $E\left( {\hat{\gamma }} \right)=10{{\log }_{10}}\frac{A_{1}^{2}A_{2}^{2}}{{{10}^{-\gamma /10}}A_{1}^{2}A_{2}^{2}}=\gamma$, which is unbiased. Therefore, variance based estimator can better cope with multipath. To verify the fact, we collect RSS measurements of 120 links on different channels in an office environment and obtain the attenuation estimation according to (8). Several links among the 120 links are obstructed by the target. The attenuation estimation results for two estimators are shown in Fig.3. We can see attenuation obtained by variance based estimator can better reflect the true attenuation of LOS path, which is usually observed between 5dB to 10dB. However, the most attenuation obtain by mean based estimator is lower than 4dB.  Hence we use variance based estimator in the rest of the paper.

\begin{figure}[!t]
\centering
\includegraphics[width=3in]{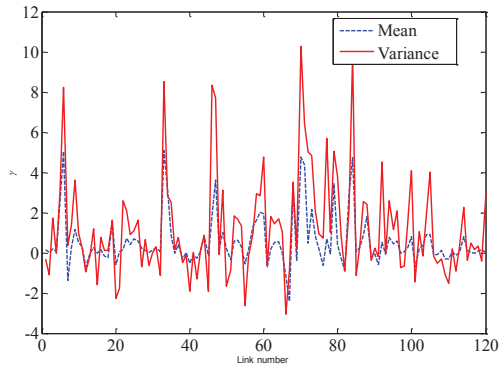}
\caption{ Attenuation estimation using mean and variance based estimator.}
\label{fig_sim}
\end{figure}

We can see from (8), when the LOS path is not affected,$\hat{\gamma }$ will be 0 and when the LOS path is obstructed $\hat{\gamma }$ will be above  0, if there are enough channels.  However, for single channel, the phase of each path is fixed and the variation of power of received signal is
\begin{equation}
\begin{aligned}
  & {{P}_{1}}-{{P}_{2}}={{\left| \underset{i=1}{\overset{P}{\mathop \sum }}\,{{A}_{i}}{{e}^{j{{\theta }_{i}}}} \right|}^{2}}-{{\left| {{10}^{-\gamma /20}}{{A}_{1}}+\underset{i=2}{\overset{P}{\mathop \sum }}\,{{A}_{2}}{{e}^{j\left( {{\theta }_{i}}-{{\theta }_{1}} \right)}} \right|}^{2}} \\
 & =\left( 1-{{10}^{-\gamma /10}}A_{1}^{2} \right)+2\left( 1-{{10}^{-\gamma /20}} \right){{A}_{1}}{{A}_{M}}\cos {{\theta }_{M}} \\
 & =\left( 1-{{10}^{-\gamma /20}}{{A}_{1}} \right)\left[ \left( 1+{{10}^{-\gamma /20}} \right){{A}_{1}}+2{{A}_{M}}\cos {{\theta }_{M}} \right] \\
\end{aligned}
\end{equation}
where $\sum\nolimits_{i=2}^{P}{{{A}_{i}}{{e}^{j\left( {{\theta }_{i}}-{{\theta }_{1}} \right)}}}={{A}_{M}}{{e}^{j{{\theta }_{M}}}}$ is the phasor sum of the multipath signals. From (9), we can see the change of power closely depends on the phase ${{\theta }_{M}}$. If ${{A}_{1}}>-\frac{2{{A}_{M}}\cos {{\theta }_{M}}}{1+{{10}^{-\gamma /20}}}$, the power of received signal will be attenuated when the link is obstructed. Otherwise the power will increase. Therefore, obstructed detection using single channel is only appropriate for the case that the power of LOS path is dominant. In the dense multipath environments, obstructed link detection using single channel cannot work well.In summary, multichannel obstructed link detection can eliminate the influence of phase and make the link detection more accurate.

It's mentioned that the power of LOS path is severely attenuated when blocked by target. Then a link is detected to be obstructed link if
\begin{equation}
{{\hat{\gamma }}_{l}}=\frac{10{{\log }_{10}}\sum\nolimits_{c=1}^{C}{{{\left[ {{P}_{c,l}}-\left( \frac{1}{C}\sum\nolimits_{c=1}^{C}{{{P}_{c,l}}} \right) \right]}^{2}}}}{10{{\log }_{10}}\sum\nolimits_{c=1}^{C}{{{\left[ {{{\bar{P}}}_{c,l}}-\left( \frac{1}{C}\sum\nolimits_{c=1}^{C}{{{{\bar{P}}}_{c,l}}} \right) \right]}^{2}}}}>{{\gamma }_{th}}
\end{equation}
where ${{\hat{\gamma }}_{l}}$ is attenuation estimation of link $l$ and ${{\gamma }_{th}}$ is attenuation threshold. The selection of ${{\gamma }_{th}}$ depends on the trade-off between probability of false alarm and missing and missing detection. The threshold will be treated in the experiment.

It should be noted that if the multipath signal is obstructed by the target, the $\hat{\gamma }$ will still be above 0 because the above detection method treats each path equally. Any strong path is blocked can lead significant enhancement in $\hat{\gamma }$ . Therefore, the above method cannot distinguish whether the LOS path or the strong multipath is blocked. To solve this case, we must utilize the spatial property of the attenuation.

\begin{figure}[!t]
\centering
\includegraphics[width=2.5in]{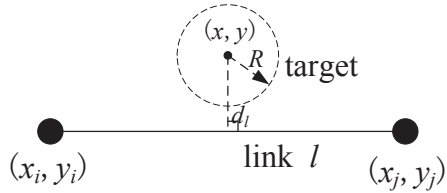}
\caption{Illustration of the spatial relationship of the target and link $l$.}
\label{fig_sim}
\end{figure}

\subsection{Spatial Property of Obstructed Links}
Next we will discuss the relationship between attenuation $\gamma$ and the position of the target. In general, the attenuation of link depends on the distance from the target to this link. The target gets closer to the link, more attenuation will be observed. On the other hand, the link will experience less attenuation. It's unlikely that the target can affect the LOS path signal when it's located far away from the link. Hence, it's reasonable to assume that the region that the target affects the LOS path is limited. Fig.4 shows the target passes through the link $l$ consisting of sensor $i$ and sensor $j$. The target is model as a cylinder with radius $R$ and the distance from the target to the link $l$ is ${{d}_{l}}$. When ${{d}_{l}}<R$, it starts to begin to affect the link. When ${{d}_{l}}>R$, the link assumed to be unaffected. Therefore, the function between attenuation $\gamma$ and distance ${{d}_{l}}$ is

\begin{equation}
{{\gamma }_{l}}=\left\{ \begin{matrix}
   f\left( {{d}_{l}},{{\varepsilon }_{l}} \right)>0,0\le {{d}_{l}}<R  \\
   f\left( {{d}_{l}},{{\varepsilon }_{l}} \right)\approx 0,{{d}_{l}}\ge R  \\
\end{matrix} \right.
\end{equation}
where ${{\varepsilon }_{l}}$ is the parameter related link $l$. The function should be also a monotonically decreasing function in the interval $\left[ 0,R \right]$. In general, it's hard to obtain the closed form of the function in cluttered environments. In [26], the author derive a nonlinear model according to the Fresnel diffraction model in the case that there is only the target between transmitter and receiver. In fact, the indoor environment may have other objects locating along the link, making it difficult to obtain an accurate mathematical model.

We can use an indicator ${{I}_{l}}$ to represent whether link $l$ is obstructed or not. Based on the spatial property, ${{I}_{l}}$ can be computed as follows:
\begin{equation}
{{I}_{l}}=\left\{ \begin{aligned}
  & 1,\ \ {{d}_{l}}<R \\
 & 0,\ \ \text{otherwise} \\
\end{aligned} \right.
\end{equation}

\section{Robust Weighted Least Square Localization Method }
In this section, we first develop the WLS localization method and then we point out that WLS is sensitive to interference links. Next, we use the spatial property of obstructed links to detect the non-obstructed links. Since the spatial detection relies on the position of the target, a coarse position estimation method which is robust to the interference links is also presented. Last, we propose the RWLS method.
\subsection{Weighted Least Square Localization Method}

 We denote ${{L}_{D}}=\left\{ l:{{{\hat{\gamma }}}_{l}}>{{\gamma }_{th}} \right\}$ is the set of obstructed links obtained from  multichannel detection and we wish to estimate the target's position from the obstructed link set. Suppose the number of links in ${{L}_{D}}$ is ${{N}_{L}}$. If link $l\in {{L}_{D}}$ constituted by sensor $i$ and sensor $j$ is obstructed, the target must satisfy the following straight line equation,
\begin{equation}
\begin{matrix}
   \frac{{{y}_{i}}-y}{{{x}_{i}}-x}=\frac{{{y}_{j}}-y}{{{x}_{j}}-x}  \\
   \left[ \left( {{y}_{j}}-{{y}_{i}} \right),\left( {{x}_{i}}-{{x}_{j}} \right) \right]\left[ \begin{matrix}
   x  \\
   y  \\
\end{matrix} \right]={{x}_{i}}{{y}_{j}}-{{x}_{j}}{{y}_{i}}  \\
\end{matrix}
\end{equation}

If we define ${{a}_{l}}={{y}_{j}}-{{y}_{i}}$, ${{b}_{l}}={{x}_{i}}-{{x}_{j}}$ and ${{e}_{l}}={{x}_{i}}{{y}_{j}}-{{x}_{j}}{{y}_{i}}$, (13) can be rewritten as
\begin{equation}
{{a}_{l}}x+{{b}_{l}}y={{e}_{l}}
\end{equation}

It's the straight line equation for link $l$ and we can get ${{N}_{L}}$ lines from link set ${{L}_{D}}$. Fig.5 shows the case when ${{N}_{L}}=3$. The position of the target $\left( x,y \right)$ is located within the area enclosed by the three lines, not exactly on the three lines. It's straightforward to use the point which has minimum squared distance to the lines as the position estimation of the target. The distance from target to the link $l$ is ${{d}_{l}}=\frac{\left| \left( {{e}_{l}}-{{a}_{l}}x-{{b}_{l}}y \right) \right|}{\sqrt{a_{l}^{2}+b_{l}^{2}}}$, then object function with respect to the position of the target is
\begin{equation}
\underset{\left( x,y \right)}{\mathop{\min }}\,\underset{l=1}{\overset{{{N}_{L}}}{\mathop \sum }}\,d_{l}^{2}=\underset{\left( x,y \right)}{\mathop{\min }}\,\underset{l=1}{\overset{{{N}_{L}}}{\mathop \sum }}\,\frac{{{\left( {{e}_{l}}-{{a}_{l}}x-{{b}_{l}}y \right)}^{2}}}{a_{l}^{2}+b_{l}^{2}}
\end{equation}

The above object function does not take the attenuation into consideration. We know a link subjects to more attenuation indicates that the less distance to this link. Therefore, the square distance of the link can be weighted by the $\hat{\gamma }_{l}^{2}$. Then the object function can be rewritten as

\begin{equation}
\underset{\left( x,y \right)}{\mathop{\min }}\,\underset{l=1}{\overset{{{N}_{L}}}{\mathop \sum }}\,\hat{\gamma }_{l}^{2}\frac{{{\left( {{e}_{l}}-{{a}_{l}}x-{{b}_{l}}y \right)}^{2}}}{a_{l}^{2}+b_{l}^{2}}
\end{equation}

The solution of the objection function can be called weighted least square (WLS) estimation of the target's position. To solve the object function, it can also be expressed by a more compact form as

\begin{equation}
\underset{\left( x,y \right)}{\mathop{\min }}\,{{\left( \mathbf{e}-\left[ \begin{matrix}
   \mathbf{a} & \mathbf{b}  \\
\end{matrix} \right]\left[ \begin{matrix}
   x  \\
   y  \\
\end{matrix} \right] \right)}^{T}}\mathbf{\gamma }\left( \mathbf{e}-\left[ \begin{matrix}
   \mathbf{a} & \mathbf{b}  \\
\end{matrix} \right]\left[ \begin{matrix}
   x  \\
   y  \\
\end{matrix} \right] \right)
\end{equation}
where $\mathbf{\gamma }=diag\left( ^{\hat{\gamma }_{1}^{2}}{{/}_{a_{1}^{2}+b_{1}^{2}}}{{,}^{\hat{\gamma }_{2}^{2}}}{{/}_{a_{2}^{2}+b_{2}^{2}}},\ldots {{,}^{\hat{\gamma }_{{{N}_{L}}}^{2}}}{{/}_{a_{l}^{2}+b_{l}^{2}}} \right)$, $\mathbf{a}={{\left[ {{a}_{1}},{{a}_{2}},\ldots,{{a}_{{{N}_{L}}}} \right]}^{T}}$ $\mathbf{b}={{\left[ {{b}_{1}},{{b}_{2}},\ldots ,{{b}_{{{N}_{L}}}} \right]}^{T}}$ and $\mathbf{e}={{\left[ {{e}_{1}},{{e}_{2}},\ldots ,{{e}_{{{N}_{L}}}} \right]}^{T}}$.

Taking derivative of the objection function with respect to ${{\left[ x,y \right]}^{T}}$and setting it to 0, we can get the solution as
\begin{equation}
{{\left[ \begin{matrix}
   {\hat{x}}  \\
   {\hat{y}}  \\
\end{matrix} \right]}_{WLS}}={{\left( {{\mathbf{H}}^{T}}\mathbf{\gamma H} \right)}^{-1}}{{\mathbf{H}}^{T}}\mathbf{\gamma e}
\end{equation}
where $\mathbf{H}=\left[ \mathbf{a},\mathbf{b} \right]$.

  Compared to particle filtering method which is the computation extensive, the computation burden of WLS is very small. Moreover, WLS does not rely on any measurement model which is difficult to obtain.

It's natural to assume that the accuracy of the position estimator depends on the number of links in set ${{L}_{D}}$. But it's true only when the links in set ${{L}_{D}}$ are all obstructed links. We should consider the case when the set ${{L}_{D}}$ contains non-obstructed links to investigate the robustness of the localization algorithm. Non-obstructed links appear in ${{L}_{D}}$ is common because two reasons. First, setting threshold cannot avoid the instance of false alarm. A reasonable choice of threshold should let false alarm keep a low value rather than 0. Second, as we have mentioned, the detection method is ineffective to the case that the multipath is obstructed.

Now we add a non-obstructed link in Fig.5, as shown in red line. It's far away from the position of the target and we assume ${{d}_{4}}>{{d}_{l}},l=1,2,3$. If the target's position estimation $\left( \hat{x}',\hat{y}' \right)$ minimize the object function $\sum\nolimits_{l=1}^{3}{d_{l}^{2}}$, we see the distance from $\left( \hat{x}',\hat{y}' \right)$ to link 4 is very large, which could not minimize the new object function$\sum\nolimits_{l=1}^{4}{d_{l}^{2}}$. Instead, the new position estimation $\left( \hat{x}'',\hat{y}'' \right)$ will move closer to the non-obstructed link, causing great localization error. Hence, the WLS algorithm is not robust to non-obstructed links, especially the links far away from the target.

From the spatial property (11), we know that the distance from the target to the non-obstructed links is larger than $R$ . Hence we can use spatial property of links to detect the non-obstructed link. However, the target's position is unknown in priori and the WLS localization algorithm is robust to this kind of links. Thus we cannot use the spatial property directly.
  Our idea is that we need to find a localization method which is not sensitive to non-obstructed links to get the coarse position estimation about the target and then use the spatial property to eliminate the non-obstructed links. Finally we adopt WLS algorithm to improve the localization accuracy. We refer to this localization method as robust WLS (RWLS) method.

\begin{figure}[!t]
\centering
\includegraphics[width=2.5in]{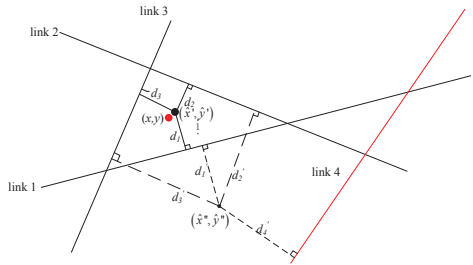}
\caption{ Illustration of WLS localization method.}
\label{fig_sim}
\end{figure}

\subsection{coarse estimation of the target}

\begin{figure}[!t]
\centering
\includegraphics[width=2.5in]{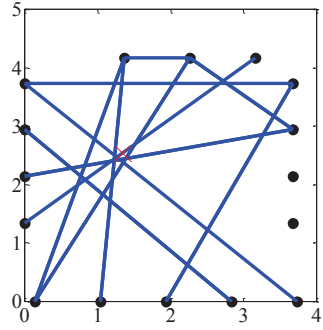}
\caption{ Obstructed links detected by multichannel RSS at real environments.}
\label{fig_sim}
\end{figure}

Fig.6 shows the obstructed links detected by multichannel RSS collected in real environments.  We see the detected links contain some non-obstructed links. However, compare to the non-obstructed link which is almost randomly distributed among the monitored area, the obstructed links almost intersect a point which is the position of the target marked by cross in the Fig.6. Therefore, although the existence of interference links, we can still infer the position of the target, which is the region most links travels cross. Hence this localization method is robust to the non-obstructed links.

\begin{figure}[!t]
\centering
\includegraphics[width=2in]{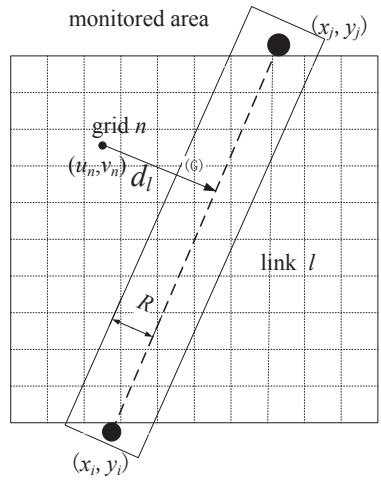}
\caption{Illustration of coarse position estimation of target.}
\label{fig_sim}
\end{figure}
Inspired by that, we first divided into the monitored area into grids and compute the number of links that passes through for each grid, as illustrated in Fig.7. The most frequently traveled grid can be seen as the coarse estimation of the target. Suppose the number of grids within the monitored area is $N={{N}_{1}}{{N}_{2}}$ , where ${{N}_{1}}$ is the number of grids on each row and ${{N}_{2}}$ is the number of grids on each column. The size of grid is $\Delta $.We denote ${{T}_{n,l}}$ is the indicator whether the link $l$travels across grid$n$and $\left( {{u}_{n}},{{v}_{n}} \right)$is the center of the grid $n$. Then the relationship between ${{T}_{n,l}}$ and $\left( {{u}_{n}},{{v}_{n}} \right)$ is
\begin{equation}
{{T}_{n,l}}=\left\{ \begin{matrix}
   1,{{d}_{n,l}}<R  \\
   0,otherwise  \\
\end{matrix} \right.
\end{equation}
where ${{d}_{n,l}}=\frac{\left| \left( {{e}_{l}}-{{a}_{l}}{{u}_{n}}-{{b}_{l}}{{v}_{n}} \right) \right|}{\sqrt{a_{l}^{2}+b_{l}^{2}}}$ is the distance from the grid $n$ to the link $l$. The frequency that grind $n$ is traveled across by the links in ${{L}_{D}}$ is

\begin{equation}
{{M}_{n}}=\sum\limits_{l=1}^{{{N}_{D}}}{{{T}_{n,l}}}
\end{equation}

Considering the attenuation of links is different, the frequency ${{M}_{n}}$ can be  weighted by
\begin{equation}
{{M}_{n}}=\sum\limits_{l=1}^{{{N}_{D}}}{{{{\hat{\gamma }}}_{l}}{{T}_{n,l}}}
\end{equation}

 The grid with highest frequency can be chosen as the coarse estimation of target's position, which can be written as
\begin{equation}
\begin{aligned}
  & {{n}_{\max }}=\arg \underset{n}{\mathop{\max }}\,{{M}_{n}} \\
 & \left( {{{\hat{x}}}_{cor}},{{{\hat{y}}}_{cor}} \right)=\left( {{u}_{{{n}_{\max }},}}{{v}_{{{n}_{\max }}}} \right) \\
\end{aligned}
\end{equation}

\subsection{Spatial detection}
After getting the coarse estimation of the target, we can use spatial property to detect the non-obstructed links. From (12), we can see the spatial detection is very simple which only requires comparing the distance from the target to the link and the radius of the target. Thus, the detection of link $l\in {{L}_{D}}$ is
\begin{equation}
{{\hat{d}}_{l}}=\frac{\left| \left( {{c}_{l}}-{{a}_{l}}{{{\hat{x}}}_{cor}}-{{b}_{l}}{{{\hat{y}}}_{cor}} \right) \right|}{\sqrt{a_{l}^{2}+b_{l}^{2}}}<{{R}_{th}}
\end{equation}
where ${{\hat{d}}_{l}}$ is the distance estimation using the coarse estimation about the target and ${{R}_{th}}$. Since the coarse estimation cannot be very accurate, ${{R}_{th}}$should be small larger than $R$.
\subsection{Robust Weighted Least Square Localization Method}
Then the entire procedure of RWLS algorithm consists of the following 4 steps: (1) Detect the obstructed links set ${{L}_{D}}$ based on the multichannel RSS variance method. (2) Obtain the coarse estimation about the target $\left( {{{\hat{x}}}_{cor}},{{{\hat{y}}}_{cor}} \right)$. (3) Detect the non-obstructed links using the spatial property and we can get a new obstructed link set ${{{L}'}_{D}}=\left\{ l:{{{\hat{d}}}_{l}}<{{R}_{th}},l\in {{L}_{D}} \right\}$. (4) Refine the position estimation using WLS method based on the link set ${{{L}'}_{D}}$.

And the detection result of link $l$ after using spatial property is
\begin{equation}
{{\hat{I}}_{l}}=\left\{ \begin{aligned}
  & 1,l\in {{{{L}'}}_{D}} \\
 & 0,\text{otherwise} \\
\end{aligned} \right.
\end{equation}

\section{Experiment Setup}
In this section, we describe the experiment setup. Section 4.A describes the sensors we use in the experiment. Section 4.B presents the description of communication protocol designed of multichannel RSS measurement. Section 4.c describes the experiment environment.
\subsection{Hardware}
The sensors employed in the experiment is TI 2530 nodes with maximum transmitting power 4.5dBm. The antenna installed on the senor is omnidirectional. Thus that when one sensor transmits signals, all other sensors can receive signals. TI2530 nodes are entirely compatible with IEEE802.15.4 protocol and can operate on the 16 channels numbered from 11 to 26 specified by IEEE 802.15.4 protocol. The sensors can provide quantized RSS value with range from 0 to 255. Because the unit of  ${{P}_{l,c}}$ is mW, the quantized RSS value should be converted from dB to mW.
\subsection{Measurement Protocol}
In order to quickly get the measurement of RSS of all links on the different channels, we have to design a communication protocol to allow the sensors work in an effective manner. The basic idea of our protocol is that first all the sensor works in one channel to get the RSS of all links on this channel and then all the nodes switch to the next channel simultaneously. In the same channel, the sensors broadcast signals in turn and when one sensor transmits signals, other sensor receive the signal and get the corresponding RSS. Thus that, the RSS of all the links in one channel can be obtained after the all the nodes have broadcasted the signals once. Then all the sensors are notified to switch to the next channel and repeat the procedure of RSS measurement in the last channel. To ensure the nodes can work in such a way, the transmitted frame can be designed as Fig.8.

The frame consists of four parts including FLAG, CID, NID and DATA. The each length of first three parts is one byte. FLAG identifies whether the frame is a data frame or a command frame. FLAG=0 means it is a data frame and vise visa. When a sensor receives a data frame, it measures the RSS of the received signal. In contrast, when the sensor receives a command frame, it switches to the channel CID. Hence CID is the channel number of current channel when FLAG=0 and the channel number of the sensor should switches to when FLAG=1. NID is the number of sensor transmitting signal. Each sensor is assigned with a unique ID in a prior. The sensor compares the NID in the received frame and ID itself to decide whether it is turn to transmit. DATA are the RSS measurements of the links the sensor connects to other sensors. Then DATA are $\left( K-1 \right)$ bytes when FLAG=0 and null when FLAG=1. The sensor with ID 1 is in charge of transmitting command frame. When the RSS measurement on the channel is over, the 1 sensor broadcasts the command frame. Moreover, to avoid the interruption due to packet loss, sensor 1 has one more function to restart the network. To this end, the senor 1 maintains a timer with time out ${{T}_{out}}=10\text{ms}$. When time out occurs, sensor 1 retransmits the data frame and otherwise it resets the timers. To reduce influence of the  disturbance from the environment on RSS measurements, the RSS of a link on each channel is measured and averaged by 100 times. The flowcharts of sensor 1 and other sensors are depicted as Fig.9.

 Besides the measurement sensors, there is a base station sensor which only receives the data frames and extracts the RSS from the data frames. The base station sensor is connected to the local PC via USB port and feeds the RSS data to the PC for post-processing.

\begin{figure}[!t]
\centering
\includegraphics[width=2in]{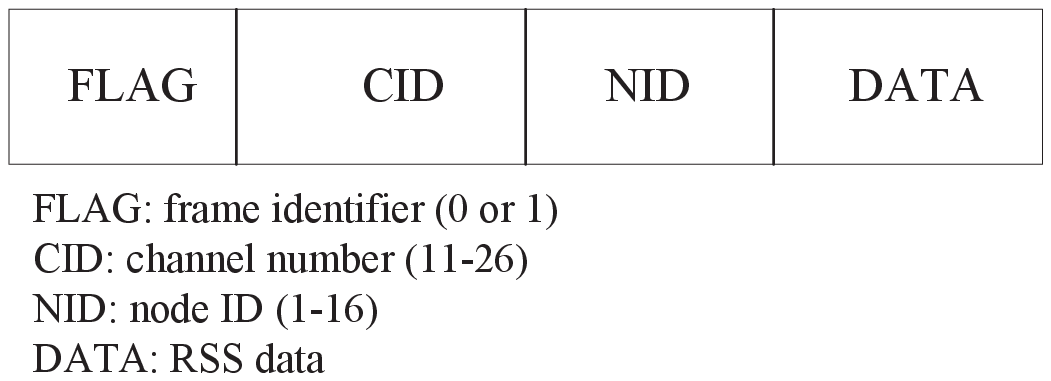}
\caption{Frame structure of the transmitted signal.}
\label{fig_sim}
\end{figure}

\begin{figure*}[ht]
\centering
\subfloat[]{\includegraphics[width=3in]{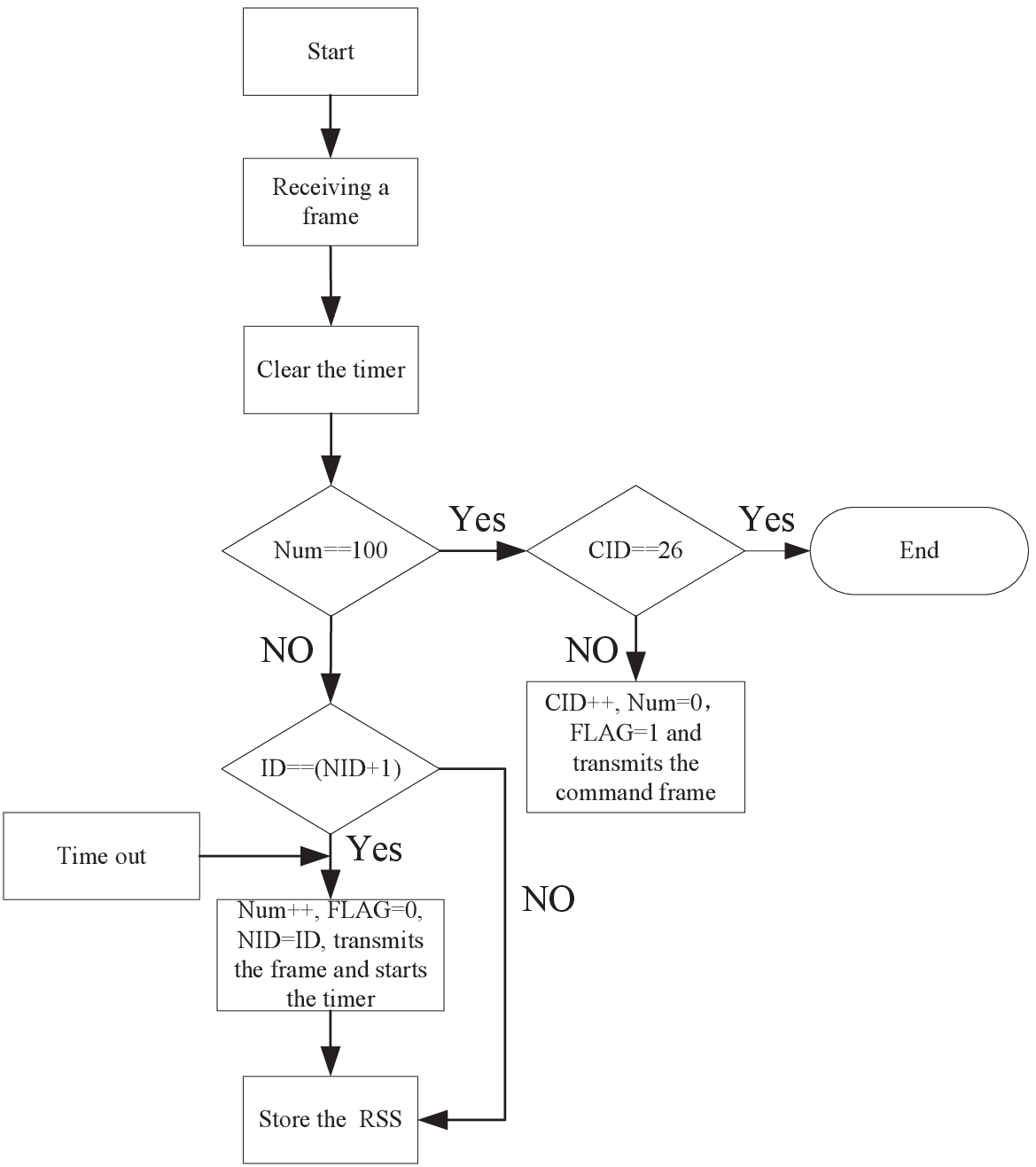}}
\hfil
\subfloat[]{\includegraphics[width=1.8in]{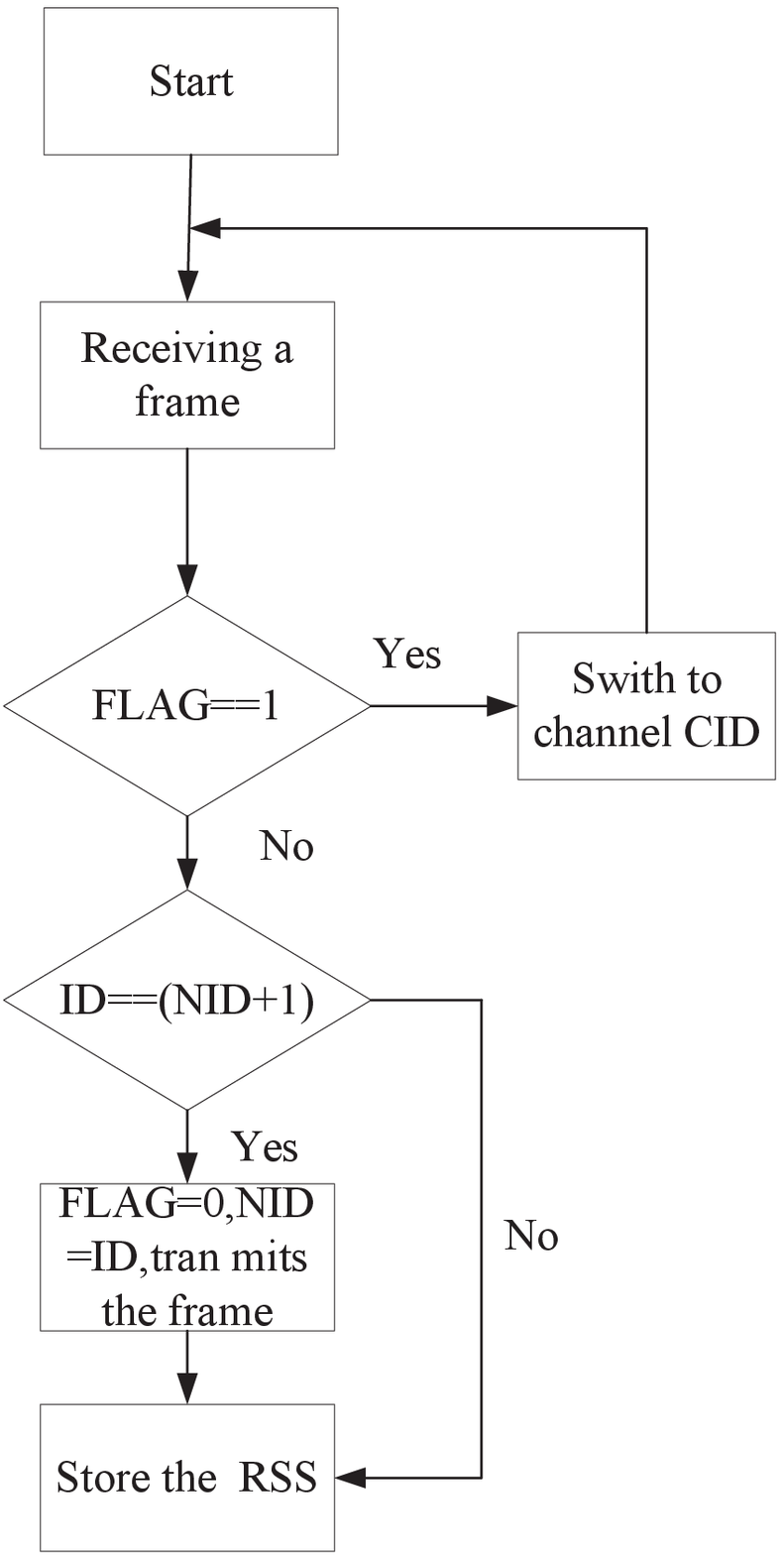}}
\caption{Flowchart of the sensors: (a) flowchart of the sensor with ID=1 and (b) flowchart of the other sensors.}
\label{fig_sim}
\end{figure*}

\subsection{Environment}
The floor plan of experiment environment with size 4.2m*3.6m is shown in Fig.10 (a) and the photography of the environment is shown in Fig.10 (b). The left side wall is made of double-layered plaster board and the bottom is a wall made of glass. The other two sides of room are brick walls. The room has also two glass windows and one wood door. In the room, there are chairs, desk, desktop, books and other stuffs. We deploy 16 sensors in the room numbered from 1 to 16 counterclockwise. The outside 9 sensors are evenly spaced with interval 0.9m on each side and the inside room 7 sensors evenly spaced with interval 0.8m on each side. Hence the signals transmitted by the outdoor sensors have to penetrate at least one wall to arrive at the sensors inside the room. All 16 sensors are fixed on the tripods with 1.2m off the ground. For evaluation of the proposed method, we choose   test positions which almost cover the entire room, as marked with crosses in Fig.10 (a). The distance of two neighbor test positions is 0.6m. A person stands at the each test position and at the same time the sensors measure the RSS and send the data to PC.

From the problem formulation, we know it's necessary to get the RSS measurements when the target is absent, which is also called calibration. The calibration can be done online [26-27] or offline [4]. In this experiment, we implement offline calibration by recoding the RSS measurements when the monitored area is free of target.

\begin{figure}
\centering
\subfloat[] {\includegraphics[width=2.5in]{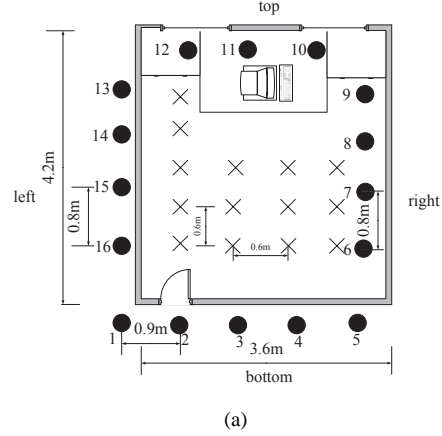}}
\\
\subfloat[] {\includegraphics[width=2.5in]{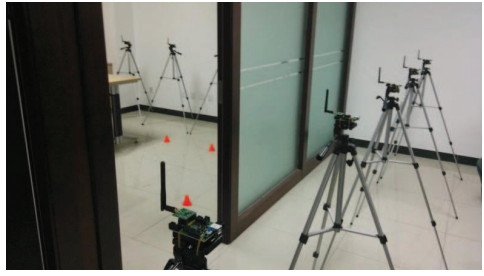}}
\caption{Experiment environment:(a) floor plan of the environment and (b) photography of the environment.}
\label{fig}
\end{figure}

\section{Experiment Results}
\subsection{Performance Metrics}
 The most frequently used metric for detection problem is the probability of missing detection ${{P}_{MD}}$ and false alarm ${{P}_{FA}}$. In out context, the missing detection refers to the obstructed links detected to be non-obstructed link and false alarm refers to the non-obstructed links detected to be obstructed links. The ${{P}_{MD}}$ and ${{P}_{FA}}$ of the proposed detection method can be computed by
\begin{equation}
\begin{aligned}
  & {{P}_{MD}}=\frac{\sum\limits_{t=1}^{T}{\sum\limits_{l=1}^{L}{{{I}_{t,l}}\left( 1-{{{\hat{I}}}_{t,l}} \right)}}}{\sum\limits_{t=1}^{T}{\sum\limits_{l=1}^{L}{{{I}_{t,l}}}}},{{P}_{FA}}=\frac{\sum\limits_{t=1}^{T}{\sum\limits_{l=1}^{L}{\left( 1-{{I}_{t,l}} \right){{{\hat{I}}}_{t,l}}}}}{\sum\limits_{t=1}^{T}{\sum\limits_{l=1}^{L}{\left( 1-{{I}_{t,l}} \right)}}}, \\
 & l=1,2,...,L,\ \ t=1,2,...,T \\
\end{aligned}
\end{equation}
where ${{I}_{t,l}}$ is the indicator of link $l$ at position $t$ and ${{\hat{I}}_{t,l}}$ is the corresponding detection result.
 In the context of DFP, the most popular evaluation metric is the root mean square error (RMSE). After obtaining the position estimation on each test position, RMSE can be calculated as
\begin{equation}
RMSE=\sqrt{\frac{1}{T}\sum\limits_{t=1}^{T}{\left[ {{\left( {{{\hat{x}}}^{t}}-{{x}^{t}} \right)}^{2}}+{{\left( {{{\hat{y}}}^{t}}-{{y}^{t}} \right)}^{2}} \right]}}
\end{equation}
where $\left( {{x}^{t}},{{y}^{t}} \right)$ is the coordinate of the target at $t$ test position and $\left( {{{\hat{x}}}^{t}},{{{\hat{y}}}^{t}} \right)$ is the position estimation.
 Another metric to measure localization accuracy is the cumulative distribution function (CDF) of the localization error, which shows the statistical property of the localization error.
\subsection{Performance Evaluation }
\subsubsection{Detection Accuracy}
Fig.11 shows the probability missing detection and false alarm versus threshold ${{\gamma }_{th}}$ for both single and multichannel detection methods. The single channel used here is channel no.11. Note that the spatial property is not considered at present. We can see for both detection methods, as threshold increases, ${{P}_{MD}}$reduces and ${{P}_{FA}}$ grows. However, there is a significant performance gap between multichannel detection method and single channel detection method. In particular, single channel detection method shows poor ability to detect the obstructed links. For example, when ${{\gamma }_{th}}=4\text{dB}$, the ${{P}_{MD}}=46.8\%$ for multichannel detection and ${{P}_{MD}}=67.8\%$ for single channel detection. The accuracy of ${{P}_{MD}}$ is improved by 40\% by multichannel detection, which indicates more available obstructed link for localization. We can see more clearly from the Fig.12 which plots the obstructed links through detection at each test position. The obstructed links detected by single channel is very unsatisfactory. It is difficult for single channel detection to determine the position of the target due to lack of truly obstructed links. However, for multichannel detection, it's easy to determine the coarse position of the target.

As for false alarm probability ${{P}_{FA}}$, when the threshold is smaller than 3dB, there are no obvious difference between the two detection methods. When the threshold is larger than 3dB, the detection accuracy of single channel is small better than that of multichannel method.

\begin{figure}[!t]
\centering
\includegraphics[width=2.5in]{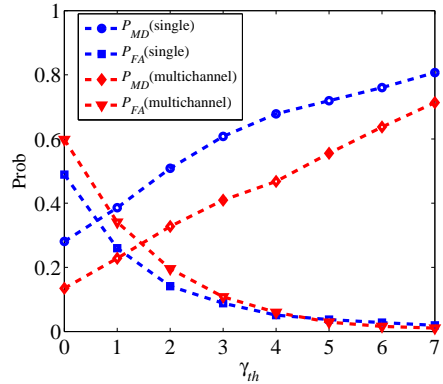}
\caption{ The probability of missing detection ${{P}_{MD}}$ and false alarm ${{P}_{FA}}$ for single channel ($c=11$) detection and multichannel detection.}
\label{fig_sim}
\end{figure}

\begin{figure}
\centering
\subfloat[] {\includegraphics[width=3.5in]{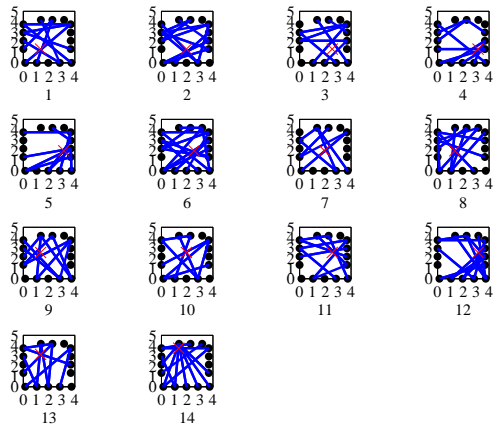}}
\\
\subfloat[] {\includegraphics[width=3.5in]{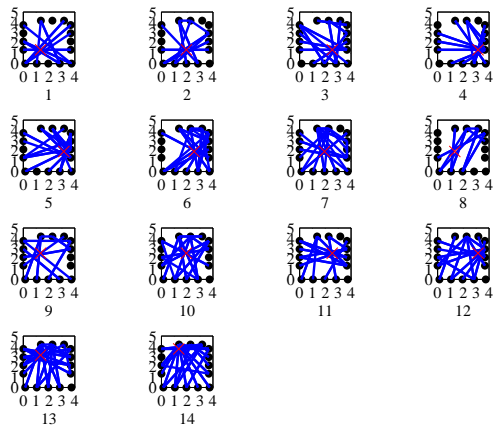}}
\caption{ Detected obstructed links for the test positions using the two detection methods when ${{\gamma }_{th}}=4\text{dB}$: (a) single channel and (b) multichannel}
\label{fig}
\end{figure}

We know before employing spatial property, it's required to get the coarse position estimation of the target. Therefore, we have to evaluate the accuracy of the coarse estimation method. Fig.13 is the box and whisker plot of localization error for coarse estimation method versus threshold. The size of grid is chosen as $\Delta =0.1\text{m}$. We can see as threshold increases, the localization error reduces rapidly. But when ${{\gamma }_{th}}$ exceeds 4dB, outliers begin to appear in this plot, meaning that there are large localization bias at some test positions. Fig.14 shows the coarse estimation result method, when ${{\gamma }_{th}}$ is 4dB, which are shown as images. The gray level of a pixel in the image represents the ${{M}_{n}}$. The brightest pixel in the image can be seen as the position estimation of the target. We see that although the existence of non-obstructed links, the coarse method still shows good localization performance.

\begin{figure}[!t]
\centering
\includegraphics[width=2.5in]{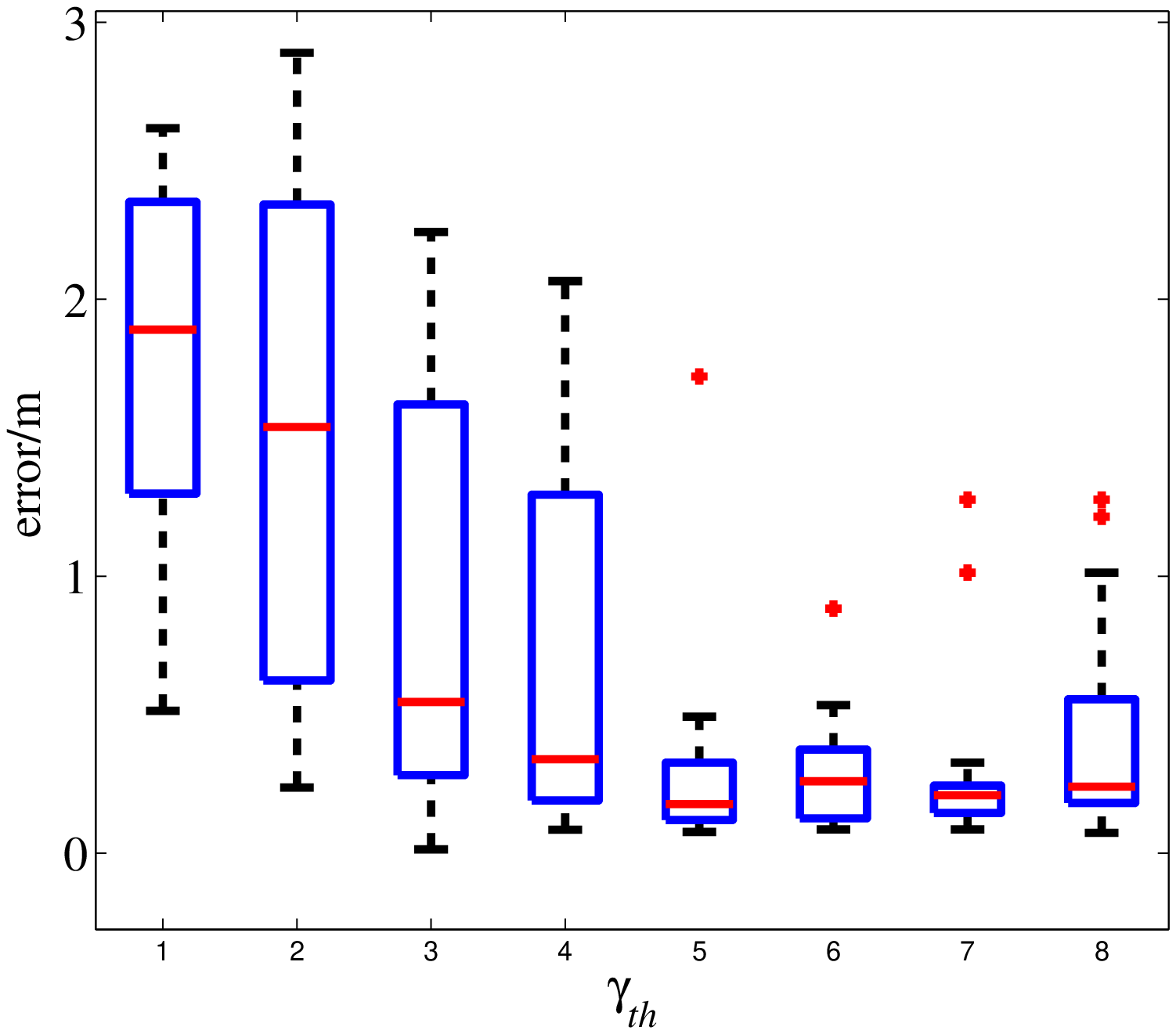}
\caption{ Box and whisker plot of localization error for coarse position estimation method.}
\label{fig_sim}
\end{figure}
\begin{figure}[!t]
\centering
\includegraphics[width=3.5in]{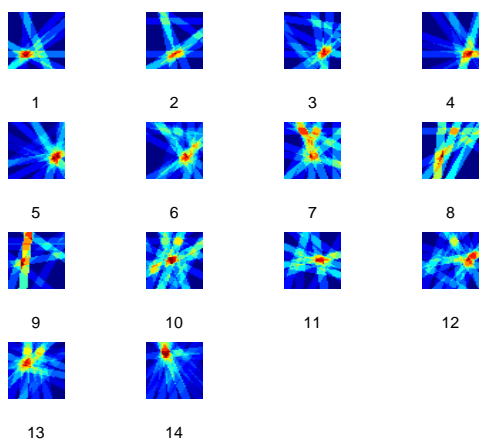}
\caption{Visual representation of the coarse estimation results.}
\label{fig_sim}
\end{figure}

After obtaining the coarse position estimation, we can use the spatial property to detect the non-obstructed links. Fig.15 shows the comparison of the detection result before and after using the spatial property. The threshold ${{R}_{th}}$is chosen as 0.5m, a smaller larger than the body radius of the target. We can see the false alarm is greatly dropped after the spatial property is employed. For example, when ${{\gamma }_{th}}$ is 4dB, the ${{P}_{FA}}$ for both are 6\% and $2\%$ respectively. As threshold increases further, the gap between the false alarm becomes larger. However, the detection performance degrades because the outliers in Fig.12 which increases the missing detection. Hence, it's appropriate to choose the threshold as 4dB. Fig.16 plots the obstructed links after spatial detection. There are almost no non-obstructed links compared to the links in Fig.12(b).

\begin{figure}[!t]
\centering
\includegraphics[width=4.0in]{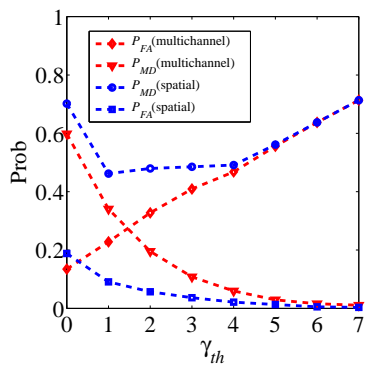}
\caption{ The probability of missing detection ${{P}_{MD}}$ and false alarm ${{P}_{FA}}$ using spatial property.}
\label{fig_sim}
\end{figure}
\begin{figure}[!t]
\centering
\includegraphics[width=3.5in]{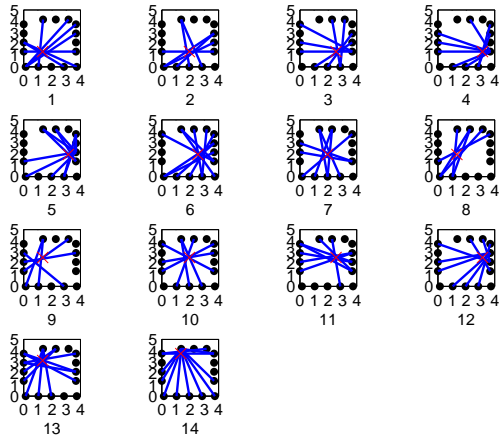}
\caption{Obstructed links detection using spatial property. }
\label{fig_sim}
\end{figure}

\subsubsection{Localization Accuracy}
Fig.17 demonstrates the localization result for WLS and RWLS methods. We see the localization error of WLS is very large. Some of the estimated positions deviate the true positions of the target. This occurs due to the fact WLS method is sensitive to the non-obstructed links. We can observe that at the test positions which have large position error for WLS method there are always some non-obstructed links which are far away from the true target position. However, after spatial detection, most of the non-obstructed links are eliminated. Therefore, the localization accuracy is greatly enhanced when adopting RWLS method. It's clear that there is only a small bias between the localization estimation for RWLS method and the true position. The RMSE of localization error for WLS and RWLS is 0.71m and 0.19m. The localization accuracy is improved by 73\%. The CDF of localization error is shown in Fig.18. We see the localization error for WLS method ranges from 0.2m to 1.6m. However, the range of localization error for RWLS method is narrowed by 0m to 0.27m.

\begin{figure}
\centering
\subfloat[] {\includegraphics[width=4in]{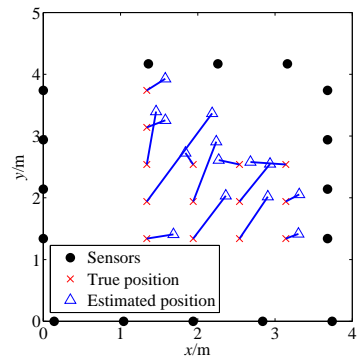}}
\\
\subfloat[] {\includegraphics[width=4in]{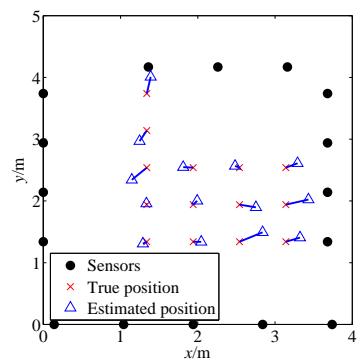}}
\caption{Scatterplot of the localization results for WLS and RWLS: (a)WLS and (b) RWLS.}
\label{fig}
\end{figure}

\begin{figure}[!t]
\centering
\includegraphics[width=3in]{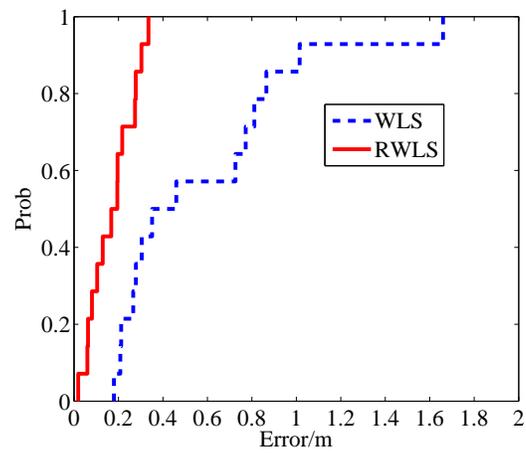}
\caption{ CDF of localization error for WLS and RWLS.}
\label{fig_sim}
\end{figure}

\subsubsection{Number of Channels}
In the previous results, maximum number of channels $C=16$ is assumed, meaning all the channels are used. Usually the estimation of attenuation of LOS path would be more accurate if more channels are available. But measuring the RSS of more channels requires more work and time. Hence we should make a balance between performance and cost. Fig.19 and Fig.20 gives the performance of detection and localization versus channel $C$ respectively.

We see as the number of channels increases, both ${{P}_{MD}}$ and ${{P}_{FA}}$ reduce. But when $C$ is larger than 8, the variation of ${{P}_{MD}}$ and ${{P}_{FA}}$ is flatter. From Fig.19 we can also observe that when $C$ is larger than 8, the localization error almost keep unchanged, if the outliers are not excluded. The difference is that as $C$ grows, the number of outliers decreases and when $C$ is larger than 14, the outliers disappear. Hence we can use fewer channels if some outliers in the localization results are allowed.

\begin{figure}[!t]
\centering
\includegraphics[width=3.5in]{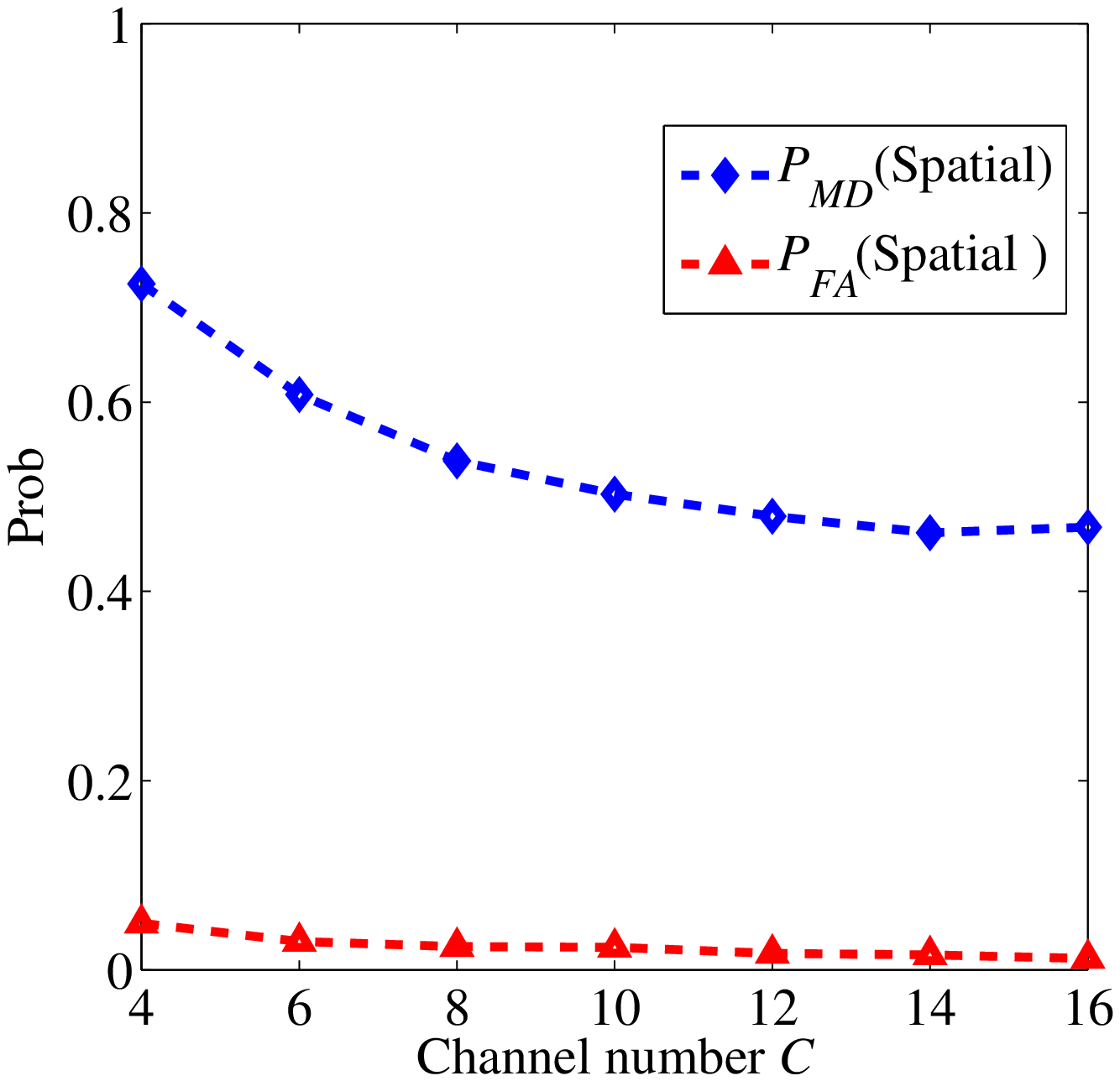}
\caption{ ${{P}_{MD}}$ and ${{P}_{FA}}$ versus number of channels $C$.}
\label{fig_sim}
\end{figure}

\begin{figure}[!t]
\centering
\includegraphics[width=3in]{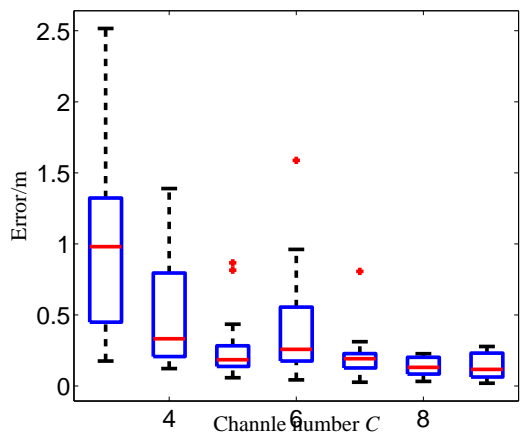}
\caption{ Box and whisker plot of localization error versus number of channels $C$.}
\label{fig_sim}
\end{figure}

\section{Conclusion}
In this paper, we develop a multichannel RSS-based DFP to improve the localization accuracy in the cluttered environments. Our method consists of two steps: obstructed links detection and localization. Due to multipath, the variation of RSS caused by the presence of target under a single channel is unpredictable even the link is obstructed. We proposed a multichannel obstructed link detection to exploit the variation of RSS on different channel, which proves to be effective in cluttered environments. Moreover, we propose a localization method termed as RWLS method which has low complexity and robust to the interference links. The experiment results conducted in a multipath rich environment show that accuracy of obstructed link detection using multichannel RSS is greatly enhanced compared to using only a single channel.  And the RMSE of localization error for RWLS is 0.19m, which is accurate enough for cluttered environments.

% or


%\appendices
%% you can choose not to have a title for an appendix
%% if you want by leaving the argument blank
%\section{}

\section*{Acknowledgment}

This work was supported in part by National Natural Science
Foundation of China (No. 61101129 and No. 61227001).

\ifCLASSOPTIONcaptionsoff
  \newpage
\fi

\end{document}